\documentclass{article}
\usepackage{amsthm,amsmath,amssymb,bm,array,chicago,epic,graphpap,graphics,float,tabularx}
\numberwithin{equation}{section}
\def\R{{\mathbb R}}
\def\RR{{\bf R}}
\def\RRbar{{\overline\RR}}
\def\GG{{\bf G}}
\def\GGbar{{\overline\GG}}
\def\mtil{{\widetilde\mu}}
\def\rhat{{\hat r}}
\def\mhat{{\widehat\mu}}
\def\bghat{{\widehat\bg}}
\def\bsihat{{\widehat\bsi}}
\def\eqdef{{\triangleq}}
\def\m{{\mu}}
\def\Xhat{{\widehat{X}}}
\def\Xtil{{\widetilde{X}}}
\def\s{{\sigma}}
\def\sumi{{\sum_{i=1}^n}}
\def\sumk{{\sum_{k=1}^n}}
\def\bsi{{\bm \sigma}}
\def\bla{{\bm \lambda}}

\def\limT{{\lim_{T\to \infty}}}
\def\lt{{\Lambda}}
\def\intT{{\int^T_0}}
\def\ph{{\phi}}
\def\bg{{\bm g}}
\def\g{{\gamma}}
\def\p{{\pi}}

\def\half{{\frac{1}{2}}}
\def\as{{\mathrm a.s.}}
\def\t{{\tau}}

\def\thhat{{\widehat \theta}}
\def\eqapprox{{\cong}}
\newcommand{\brac}[1]{\langle#1\rangle}

\begin{document}

\title{\bf A second-order stock market model }
\author{Robert Fernholz\footnote{INTECH, One Palmer Square, Princeton, NJ 08542.}  \and
Tomoyuki Ichiba\footnote{Department of Statistics and Applied Probability, South Hall, University of California, Santa Barbara, CA 93106.} \and Ioannis Karatzas\footnote{INTECH, One Palmer Square, Princeton, NJ 08542.} }
\date{February 12, 2012}
\maketitle
\begin{abstract} 
A \emph{first-order model} for a stock market  assigns to each  stock  a return parameter and a variance parameter that depend only on the \emph{rank} of the stock. A  \emph{second-order model}  assigns these parameters based on both the rank and the \emph{name} of the stock. First- and second-order models exhibit stability properties that make them appropriate as a backdrop for the analysis of the idiosyncratic behavior of individual stocks. Methods for the estimation of the parameters of second-order models are developed in this paper.
\end{abstract}

\vspace{.15in}


\section{Introduction}

First-order and second-order stock market models are relatively simple stochastic models that manifest some of the stability properties of actual stock market behavior. These models are descriptive as opposed to normative, and are constructed using data analysis based on actual stock markets. First-order models are stock-market models where the parameters for return and volatility  are based on the ranks of the stocks. These models were introduced in \citeN{F:2002} and developed in \citeN{BFK:2005}, and reflect the actual rank-based growth rates and variances of the stocks in the market. First-order models are asymptotically stable, and accurately reproduce the long-term characteristics of the market's capital distribution. However, these models are ergodic in the sense that each stock asymptotically spends equal  average time at each rank, and this ergodicity property does not seem to be present in actual markets. This lack of verisimilitude is the motivation to consider the next level of complexity: second-order models.

Second-order models are a form of hybrid Atlas models, where the return and volatility parameters are based on the rank and the name (or index) of the stocks (see \shortciteN{IPBKF:2011}). While these models retain many of the characteristics of first-order models, the above ergodicity property is no longer present, and this produces a more realistic representation of actual stock market behavior. In second-order models, larger stocks tend to remain asymptotically among larger stocks, and smaller stocks tend to remain among smaller stocks. This behavior is closer to that of actual stock markets, so second-order models provide a more accurate descriptive representation of stock market behavior. 

Estimation of the parameters for first-order models is fairly straightforward, and  can be accomplished without great ado. Second-order parameter estimation is somewhat more complicated. Here we shall focus on the growth-rate parameters, and find it necessary to rely on implicit methods to determine values for these parameters. Our purpose here is to develop techniques for estimating second-order growth-rate parameters, not to carry out an exhaustive examination of these parameters for an entire stock market. First, let us establish some formal definitions.

A \emph{market}  is a family of \emph{stocks} $X=(X_1,\ldots,X_n)$ whose capitalizations are modeled by continuous, positive semimartingales that satisfy
\begin{equation}\label{00}
d\log X_i(t) = G_i(t)\,dt + \sum_{\nu=1}^d S_{i\nu}(t)\,dB_\nu(t),
\end{equation}
for $t\in\R$, where $n\le d$, $B=(B_1,\ldots,B_d)$ is an $\R^d$-valued Brownian motion defined on $\R$, and the $G_i$ and $S_{i\nu}$ are progressively measurable with respect to the Brownian filtration, with $G_i$ locally integrable and $S_{i\nu}$ locally square-integrable. The reason we define these processes on $\R$ is that in practice we are confronted with time series over a given block of time, and the analysis of these series can be performed in both forward and reversed  time. Hence, we see a sample in time of the processes $X_1,\ldots,X_n$ and  draw our conclusions from this sample.

We shall assume that  for any $i\ne j$, the intersection sets $\{t:X_i(t)=X_j(t)\}$  have Lebesgue measure zero, almost surely, and we shall also assume that there are no {\em triple points}, i.e., if $i<j<k$ then there is almost surely no $t\in\R$ such that $X_i(t)=X_j(t)=X_k(t)$. The general setting for our model can be found in \citeN{F:2002} and \citeN{FK:2009}.

The value $X_i(t)$ of the stock $X_i$ at time $t$ represents the total capitalization of the company at that time. If we let $Z$ represent the total capitalization of the market, then
\begin{equation*}
Z(t)\eqdef X_1(t)+\cdots+X_n(t),
\end{equation*}
and we can define the \emph{market portfolio} to be the portfolio $\m$ with weight processes given by the \emph{market weights}
\begin{equation*}
\m_i(t)\eqdef \frac{X_i(t)}{Z(t)}, \quad\text{ for }i=1,\ldots,n.
\end{equation*}

We shall assume that the market
weight process $\m=(\m_1,\ldots,\m_n)$ has a stable, or \emph{steady-state,} distribution, and
that the system is in that stable distribution. We shall be interested in the relative behavior of the log-capitalizations or log-weights. If $\m(t)$ is in its steady-state distribution, then the
\emph{log-difference processes} defined by
\begin{equation*}
\log X_i(t)-\log X_j(t)=\log\m_i(t)-\log\m_j(t),
\end{equation*}
 for $i,j=1,\ldots,n$, will also be in their steady-state distribution. 

Consider the \emph{ranked capitalization processes} corresponding to the $X_i(t)$ in descending order
\begin{equation*}
X_{(1)}(t)\ge\cdots\ge X_{(n)}(t),
\end{equation*}
and the corresponding \emph{ranked market weights}
\begin{equation*}
\m_{(1)}(t)\ge\cdots\ge \m_{(n)}(t).
\end{equation*}
Let $r_t(i)$ represent the rank of $X_i(t)$, and let $p_t$ be the inverse
permutation of $r_t$ (with ties in rank settled by the order of the indices), so 
\begin{equation*}
X_i(t)=X_{(r_t(i))}(t)\quad \text{and}\quad  X_{(k)}(t)=X_{p_t(k)}(t),
\end{equation*}
and, similarly
\begin{equation*}
\m_i(t)=\m_{(r_t(i))}(t) \quad \text{and} \quad \m_{(k)}(t)=\m_{p_t(k)}(t).
\end{equation*}
Hence, $p_t(k)$ represents the index, or {\em name}, of the stock occupying rank $k$ at time $t$.

The ranked market weights $(\m_{(1)}(t),\ldots,\m_{(n)}(t))\equiv(\m_{p_t(1)}(t),\ldots,\m_{p_t(n)}(t))$ comprise the \emph{capital distribution curve} of the market at time $t$. The capital distribution curves over several decades of the 20th century can be seen in Figure~\ref{f1}, a version of which appears in \citeN{F:2002}. The curves in Figure~\ref{f1} show the ranked market weights on December 31 of the years 1929, 1939, 1949, 1959, 1969, 1979, 1989, and 1999. During that period, the number of stocks in the market increased over each decade, so the decade associated with each curve is clear from the chart. We see that the capital distribution curve of the market shows a certain stability over time, so the assumption that $\m$ is in its steady state distribution would seem to be consistent with the observed data.

\vspace{10pt}
\begin{figure}[H]
\begin{center}
\scalebox{.5}{ \rotatebox{90}{\includegraphics{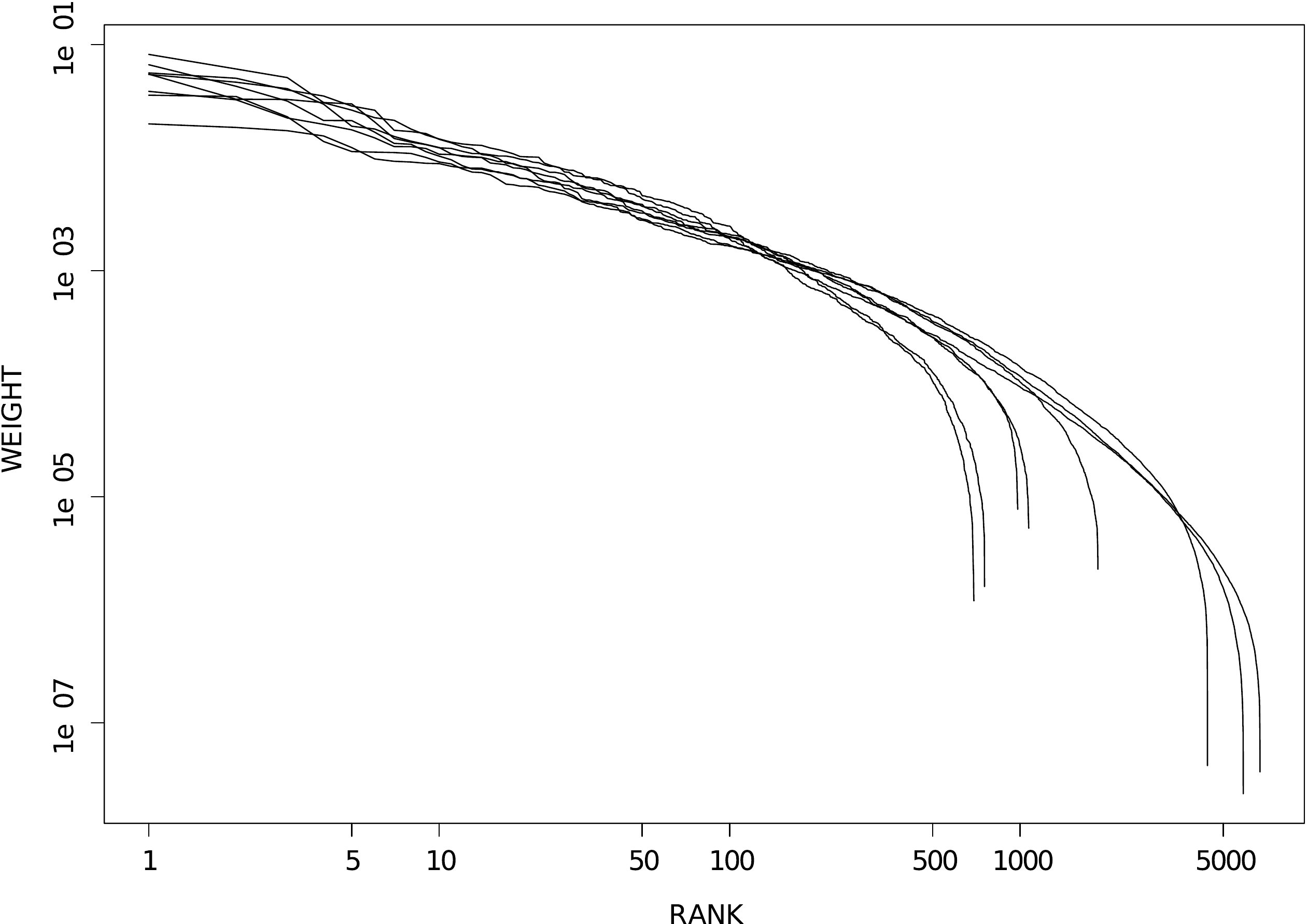}}}
\end{center}
\caption{Capital distribution of the U.S.\ market: 1929--1999.} 
\centerline{The curves show the ranked weights at the end of each decade.}
\end{figure}\label{f1}

\vspace{10pt}
\noindent{\bf\large Acknowledgements.} The authors are grateful to Adrian Banner, Daniel Fernholz, Vassilios Papathanakos, and Johannes Ruf for their many helpful discussions and suggestions, as well as for their participation and inspiration during the course of this research.

\section{First-order models}

A first-order model is a stock-market model in which each stock has constant growth and variance parameters that depend only on the {\em rank} of the stock by market capitalization. These models were developed in \citeN{F:2002} and \shortciteN{BFK:2005}, and can be constructed to reflect certain properties of actual stock markets. A first-order model that is based on an actual market will have a steady-state capital distribution curve that is about the same as the capital distribution curve for the actual market (see \citeN{F:2002}, Figure 5.6).

A \emph{first-order model}  is  defined by a system $\Xhat=(\Xhat_1,\ldots,\Xhat_n)$ of the form
\begin{align*}
d\log \Xhat_i(t) 
&= g_{\rhat_t(i)}\,dt + \s_{\rhat_t(i)}\,dW_i(t),\\
&= \sumk g_k\b1{\rhat_t(i)=k}dt 
+ \sumk\s_k\b1{\rhat_t(i)=k}dW_i(t),
\end{align*}
for $i=1,\ldots,n$, where $g_1,\ldots,g_n$ are real constants, $\s_1,\ldots,\s_n$ are positive constants, and $(W_1,\ldots,W_n)$ is an $\R^n$-valued Brownian motion, and where $\rhat_t(i)$ represents the rank of $\Xhat_i(t)$ (analogously to $r_t(i)$ for the rank of $X_i(t)$). We shall assume that the $g_k$ satisfy
\begin{equation*}
g_1+\cdots+g_n=0, 
\end{equation*}
and 
\begin{equation*}
\sum_{k=1}^m g_k<0,
\end{equation*}
for $m<n$. With these parameters, the $\Xhat_i$ form an \emph{asymptotically stable} system, which means  that the market weights $\mhat_i(t)=\Xhat_i(t)/\big(\Xhat_1(t)+\cdots+\Xhat_n(t)\big)$ satisfy
\begin{equation*}
 \lim_{t\to\infty}t^{-1}\log \mhat_i(t) = 0,\quad\text{for }i=1,\ldots,n,
\end{equation*}
and the limits corresponding to \eqref{2.2a} and \eqref{2.2z} below exist (see also \citeN{F:2002}, Definition~5.3.1).

Suppose we have a market $X$,  and suppose that its market portfolio $\m$  is in the steady-state distribution.
We define the asymptotic \emph{rank-based relative variances} for the market by
 \begin{equation}\label{2.2a}
 \bsi_k^2\eqdef \lim_{t\to\infty}t^{-1}\brac{\log \m_{(k)}}(t),
 \end{equation}
 and the asymptotic \emph{rank-based relative growth rates}  by
 \begin{equation}\label{2.8}
  \bg_k\eqdef\limT{1}\intT \sumi\b1{r_t(i)=k}d\log\m_i(t),
\end{equation}
 and suppose these limits exist almost surely. Since these parameters are based on the market weight processes $\m_i$, they 
 represent values relative to the market  portfolio $\m$.
 
 For $k<\ell$, let $\lt{k,\ell}$ be the local time of the nonnegative semimartingale
$\log(\m_{(k)}/\m_{(\ell)})\ge 0$ at the origin, and set $\lt{0,1} \equiv 0 \equiv \lt{n,n+1}$. Since we have assumed that the $X_i$ almost surely have no triple points, it follows that for $\ell>k+1$,  the  local time by $\lt{k,\ell}$ is identically zero, so here we need to consider only local times of the form $\lt{k,k+1}$, and we have
\begin{equation*}
d\log\m_{(k)}(t)=\sumi\b1{r_t(i)=k}d\log\m_i(t) +\half\,d\lt{k,k+1}(t)-\half\,d\lt{k-1,k}(t),\as
\end{equation*}
For $k=1,\ldots,n-1$, we can define the \emph{asymptotic local time}
\begin{equation}\label{2.2z}
\bla_{k,k+1} \eqdef  \lim_{t\to\infty}t^{-1}{\lt{k,k+1}(t)},
\end{equation}
which exists almost surely, and define  $\bla_{0,1}\equiv 0\equiv\bla_{n,n+1}$. It turns out that the estimation of the $\bla_{k,k+1}$ is not difficult, and the procedure is described in the appendix of \citeN{F:2002}. It can be shown (c.f. Proposition 5.3.2 in \citeN{F:2002}) that
\begin{equation}\label{2.3}
\bg_k=\half\big(\bla_{k-1,k}-\bla_{k,k+1}\big),\as
\end{equation}
holds for $k=1,\ldots,n$, and it follows that $\bg_1+\cdots+\bg_n=0$. 

\begin{figure}[H]
\begin{center}
\vspace*{-40pt}
\hspace*{-10pt}
\scalebox{.85}{ \rotatebox{0}{\includegraphics{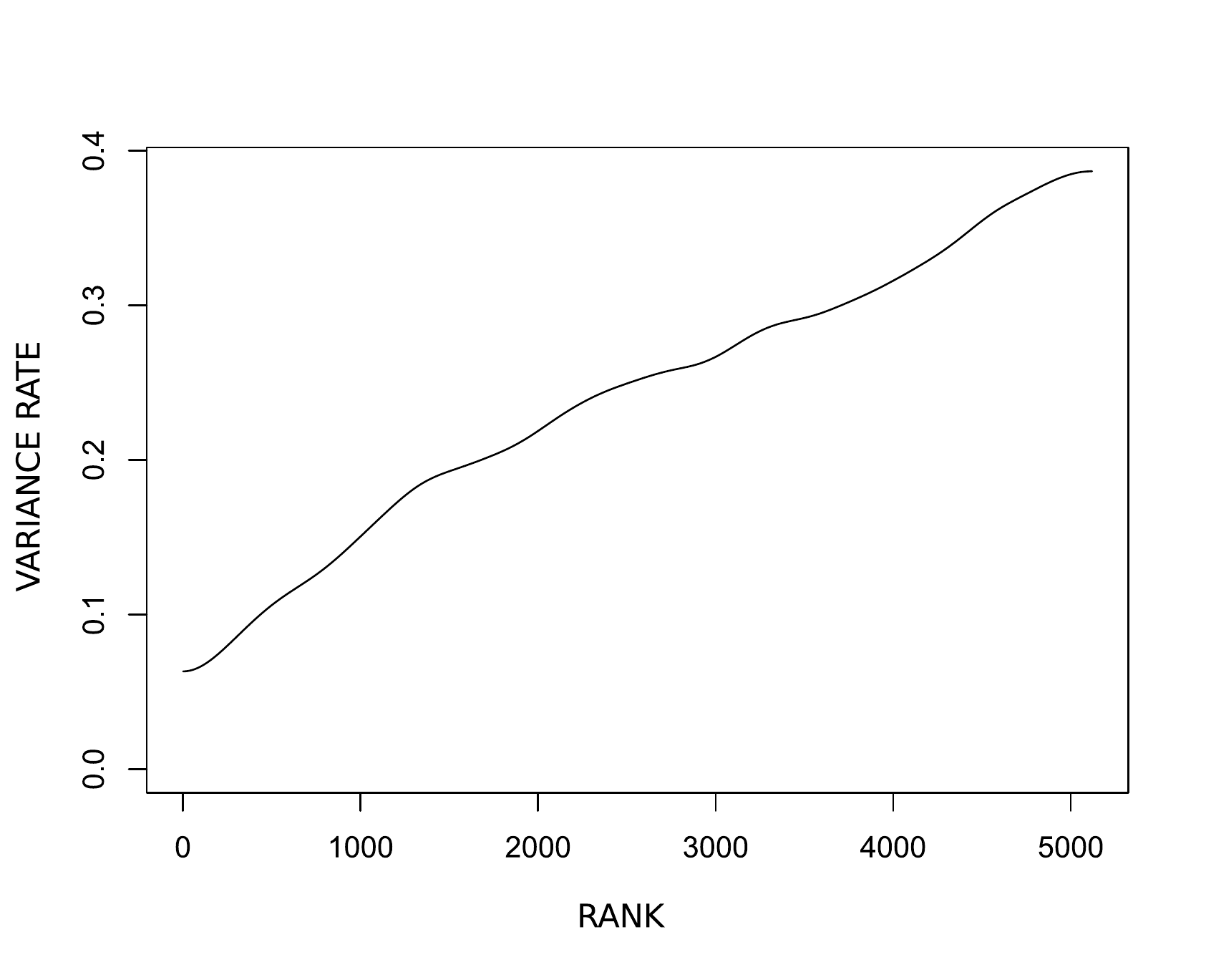}}}
\end{center}
\vspace*{-25pt}
\caption{Smoothed values of $\bsi^2_k$, $k=1,\ldots,5120$, for U.S.\ market: 1990--1999.}\label{f2}
\end{figure}

\begin{figure}[H]
\begin{center}
\vspace*{-35pt}
\hspace*{-10pt}
\scalebox{.80}{ \rotatebox{0}{\includegraphics{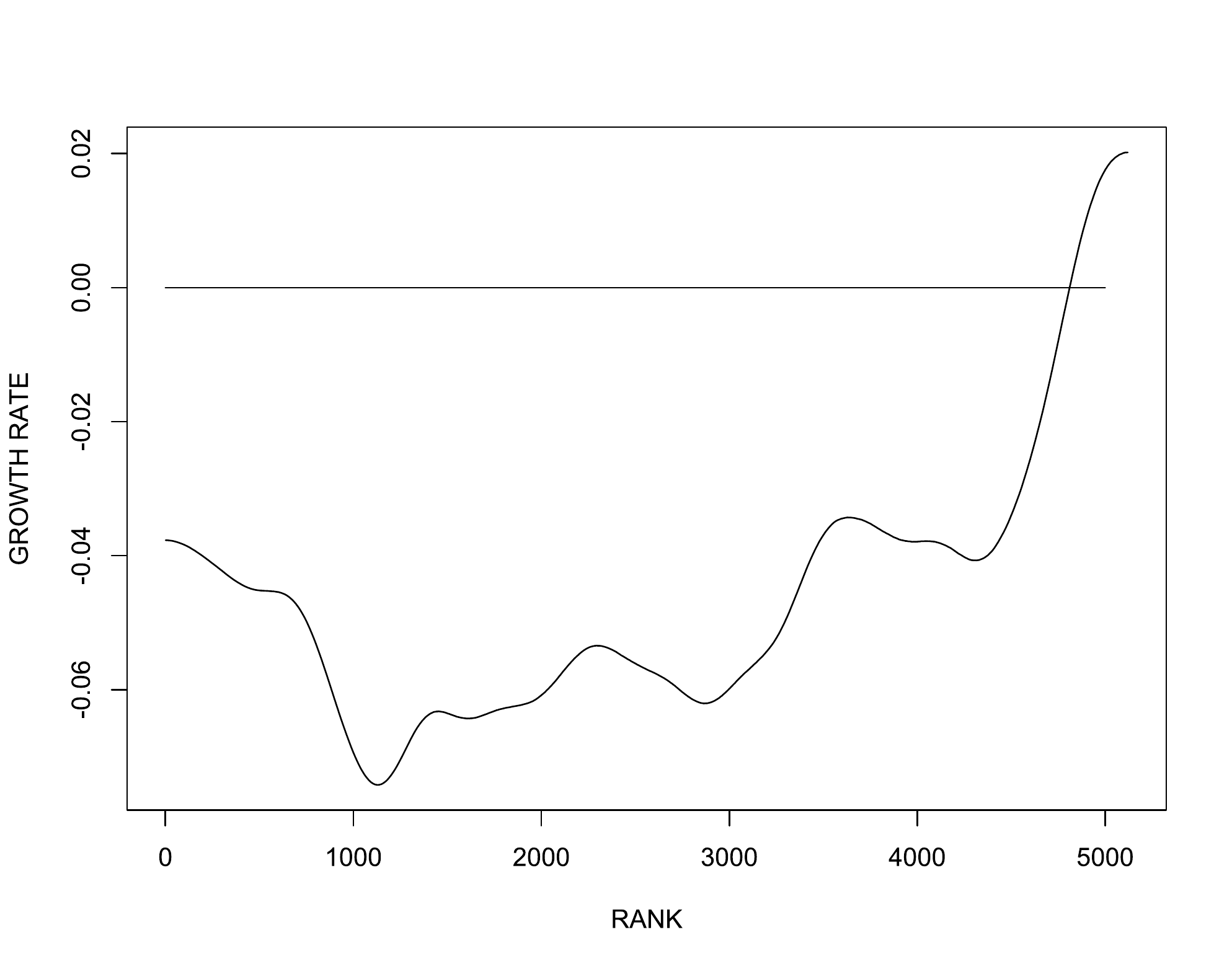}}}
\end{center}
\vspace*{-25pt}
\caption{Smoothed values of $\bg_k$, $k=1,\ldots,5120$, for U.S.\ market: 1990--1999.}\label{f3}
\centerline{The $\bg_k$ satisfy $\sumk\bg_k=0$, where $n\eqapprox 7000$.} 
\end{figure}

The smoothed values of $\bsi^2_k$ and $\bg_k$ for the largest 5120 stocks in the U.S. market for the decade 1990--1999 are shown in Figures~\ref{f2} and \ref{f3}. Since the number of stocks in the market changed during the decade of 1990--1999, we limit our attention here to the largest 5120 stocks, which is fewer than the number of stocks in the market at any time during that decade. 
The values in Figure~\ref{f3} do not add up to zero, since the largest 5120 stocks are a strict  subset of the larger market.

The first-order model $\Xhat=(\Xhat_1,\ldots,\Xhat_n)$ such that
\begin{align*}
d\log\Xhat_i(t)
&=\bg_{\rhat_t(i)}dt + \bsi_{\rhat_t(i)}dW_i(t),\\
&=\sumk\bg_k\b1{\rhat_t(i)=k}dt + \sumk\bsi_k\b1{\rhat_t(i)=k}dW_i(t),
\end{align*}
where $\rhat_t(i)$ is the rank of $\Xhat_i(t)$ at time $t$, is called the \emph{first-order model} for the market $X$. As we have seen, the growth and variance parameters for $\Xhat$ are derived from the relative growth and variance 
parameters corresponding to the market weight processes $\mhat_i$, not directly from the capitalization processes $\Xhat_i$.

\vspace{10pt}

First-order models are ergodic in the sense that
\begin{equation}\label{2.20}
\limT{1}\intT\b1{\rhat_t(i)=k}dt=\limT{1}\intT\b1{\Xhat_i(t)=\Xhat_{(k)}(t)}dt=\frac{1}{n},\as
\end{equation}
 This ergodicity property does not seem to be present in real markets, but instead there exist
\emph{asymptotic occupation rates} defined by
 \begin{equation}\label{2.2b}
\theta_{ki} \eqdef \limT{1}\intT\b1{r_t(i)=k}dt=\limT{1}\intT \b1{X_i(t)=X_{(k)}(t)}dt,\as
 \end{equation}
Here $\theta_{ki}$ represents the fraction of time that $X_i$ spends in the $k$th rank. The $n\times n$ matrix $\theta=(\theta_{ki})$ is bistochastic, and we shall assume that all the entries are positive. For a first-order model, \eqref{2.20} implies that $\theta_{ki}=1/n$ for all $i$ and $k$, and since this does not seem to characterize the behavior of real markets, we shall now consider a more general class of models.

\section{Second-order models}

A second-order model is a stock-market model in which each stock has constant growth and variance parameters that depend on the {\em rank} and {\em name}, or index, of the stock. Second-order models are examples of
\emph{hybrid (Atlas) models,} which were discussed in \shortciteN{IPBKF:2011}. A  {\em second-order model} is defined by a system $\Xhat=(\Xhat_1,\ldots,\Xhat_n)$ of the form
\begin{align}
d\log \Xhat_i(t) &= (\g_i+g_{\rhat_t(i)})dt +\s_{i,\rhat_t(i)}\,dW_i(t)\label{3.1}\\
&= \Big(\g_i+\sumk g_k\b1{\rhat_t(i)=k}\Big)dt 
+ \sumk\s_{ik}\b1{\rhat_t(i)=k}dW_i(t),\notag
\end{align}
for $i=1,\ldots,n$, with constants $g_k$, $\g_i$ and $\s_{ik}>0$, for $i,k=1,\ldots,n$,  and a Brownian motion $W$. In order for the  $\Xhat_i$ to be asymptotically stable, these parameters must satisfy
\begin{equation*}
g_1+\cdots+g_n=0=\g_1+\cdots+\g_n,
\end{equation*}
 and, for any permutation $\p\in\Sigma_n$,
\begin{equation*}
\sum_{k=1}^m(g_k+\g_{\p(k)})<0,\quad\text{for $m<n$.}
\end{equation*} 
Here we are interested in estimating the growth-rate parameters $\g_i$ and $g_k$. For simplicity, we shall consider only rank-based variances, and assume that  $\s_{ik}^2=\s^2_{k}$ for all $i$ and $k$.

It was shown in \shortciteN{IPBKF:2011} that a second-order model of the form \eqref{3.1} is asymptotically stable, and the asymptotic occupation rates
\begin{equation}\label{3.4}
\thhat_{ki}\eqdef \limT{1}\intT \b1{\rhat_t(i)=k}\,dt
\end{equation}
are defined for all $i$ and $k$, almost surely. The matrix $\thhat=(\thhat_{ki})$, like $\theta$ in \eqref{2.2b}, will be bistochastic with positive entries. We can generate the first-order parameters  $\bsihat^2_k$  and $\bghat_k$ for $\Xhat$ as in \eqref{2.2a} and \eqref{2.8}, with 
 \begin{equation*}
 \bsihat_k^2\eqdef \lim_{t\to\infty}t^{-1}\brac{\log \mhat_{(k)}}(t),
 \end{equation*}
 and  
 \begin{equation*}
  \bghat_k\eqdef\limT{1}\intT \sumi \b1{\rhat_t(i)=k}d\log\mhat_i(t).
\end{equation*}
With these parameters, it was shown in \shortciteN{IPBKF:2011} that, almost surely,
\begin{align}
\bghat_k &= g_k + \sumi\thhat_{ki}\g_i\label{3.5}\\
0&= \g_i + \sumk\thhat_{ki}g_k.\label{3.6}
\end{align}
In matrix form, this can be expressed
\begin{align*}
\bghat&=g+\thhat\g\\
0&=\g+\thhat^T g,
\end{align*}
where $\g$, $g$, and $\bghat$ are column vectors. From this we see that
\begin{equation}\label{4}
\g=-\thhat^T g,
\end{equation}
so
\begin{equation}\label{1}
\bghat = \big(I_n-\thhat\thhat^T\big)g.
\end{equation}

\section{Estimation of second-order parameters}

The  first-order growth parameters  $\bg_k$  for the market $X$ can be estimated directly from  the stock return time series; however,  second-order growth parameters will have to be estimated indirectly. We wish to construct a second-order model that has first-order growth parameters equal to those of the market, and an occupation-rate matrix equal to the occupation-rate matrix $\theta$ of the market. Under these circumstances, as in  \eqref{1}, we have
\begin{equation}\label{4.1a}
\bg = \big(I_n-\theta\theta^T\big)g,
\end{equation}
and we wish to solve this equation for  $g$, the vector of name-based growth parameters for the  second-order model of  the market $X$. If we can solve \eqref{4.1a} for this $g$, then we can use \eqref{4}, in the form
\begin{equation}\label{4.1b}
\g=-\theta^T g,
\end{equation}
to generate the name-based growth parameters $\g_i$ for this second-order model. Let us first consider the matrix $\theta$.

The matrix $\theta$ is bistochastic and we have assumed that all its entries are positive, so this also holds for $\theta^T$ and $\theta\theta^T$. By the Perron-Frobenius theorem (see \citeN{Perron:1907}), the symmetric matrix $\theta\theta^T$ will have a simple eigenvalue equal to 1 with eigenvector $e_1=(1,1,\ldots,1)'$, and all the other eigenvalues will have absolute value less than 1. Hence, $I_n-\theta\theta^T$ has rank $n-1$ and its kernel is generated by $e_1$, so the condition that the $g_k$ sum to zero means that $g$ is orthogonal to this kernel, and this ensures a unique solution to \eqref{4.1a}.

Unfortunately,  it seems to be essentially impossible to estimate $\theta$ with any reasonable accuracy, so although we can use this matrix to prove the existence and uniqueness of $g$, in practice we cannot actually solve equation \eqref{4.1a}. 
Instead, let us consider \eqref{3.5} in the form
\begin{equation}\label{4.3a}
\bg_k = g_k + \sumi\theta_{ki}\g_i.
\end{equation}
We can use this equation to generate the $g_k$ recursively, and then estimate the $\g_i$  from the returns data and the $g_k$.

Let  us assume that the market $X$ is defined for all $t\in\R$, that the weight process $\m$ for $X$ has a stable distribution, and that $\m$ is in that stable distribution. We can then define the \emph{time-reversed} market $\Xtil$ with stock capitalizations $\Xtil_i(t)\eqdef X_i(-t)$ and weights $\mtil_i(t)\eqdef \m_i(-t)$, and with this definition we can define the expected backward occupation rates similarly to \eqref{2.2b}. Since the weight process is in its steady-state distribution, the limits of \eqref{2.2b} will be the same at plus and minus infinity, so the forward and backward expected occupation rates $\theta_{ki}$ will be equal.  The results of \citeN{Bertoin:1987} imply that the forward and backward asymptotic local times $\lt{k,k+1}$ will also be the same, so the forward and backward versions of the $\bla_k$ are equal. Hence, it follows from \eqref{2.3} that the forward and backward $\bg_k$ are equal. In this case, \eqref{4.1a} implies that the forward and backward values of the $g_k$ are equal, and from  \eqref{4.1b}, we see that the forward and backward $\g_i$ are also equal. Quadratic variation is invariant under time reversal, so the forward and backward $\bsi_k$ will be the same. Hence, the  first- and second-order models for $X$ are the same as the corresponding models for $\Xtil$, and this allows us to use both $X$ and $\Xtil$ to estimate the second-order parameters.

\begin{figure}[H]
\begin{center}
\vspace*{-15pt}
\hspace*{-20pt}
\scalebox{.58}{ \rotatebox{90}{\includegraphics{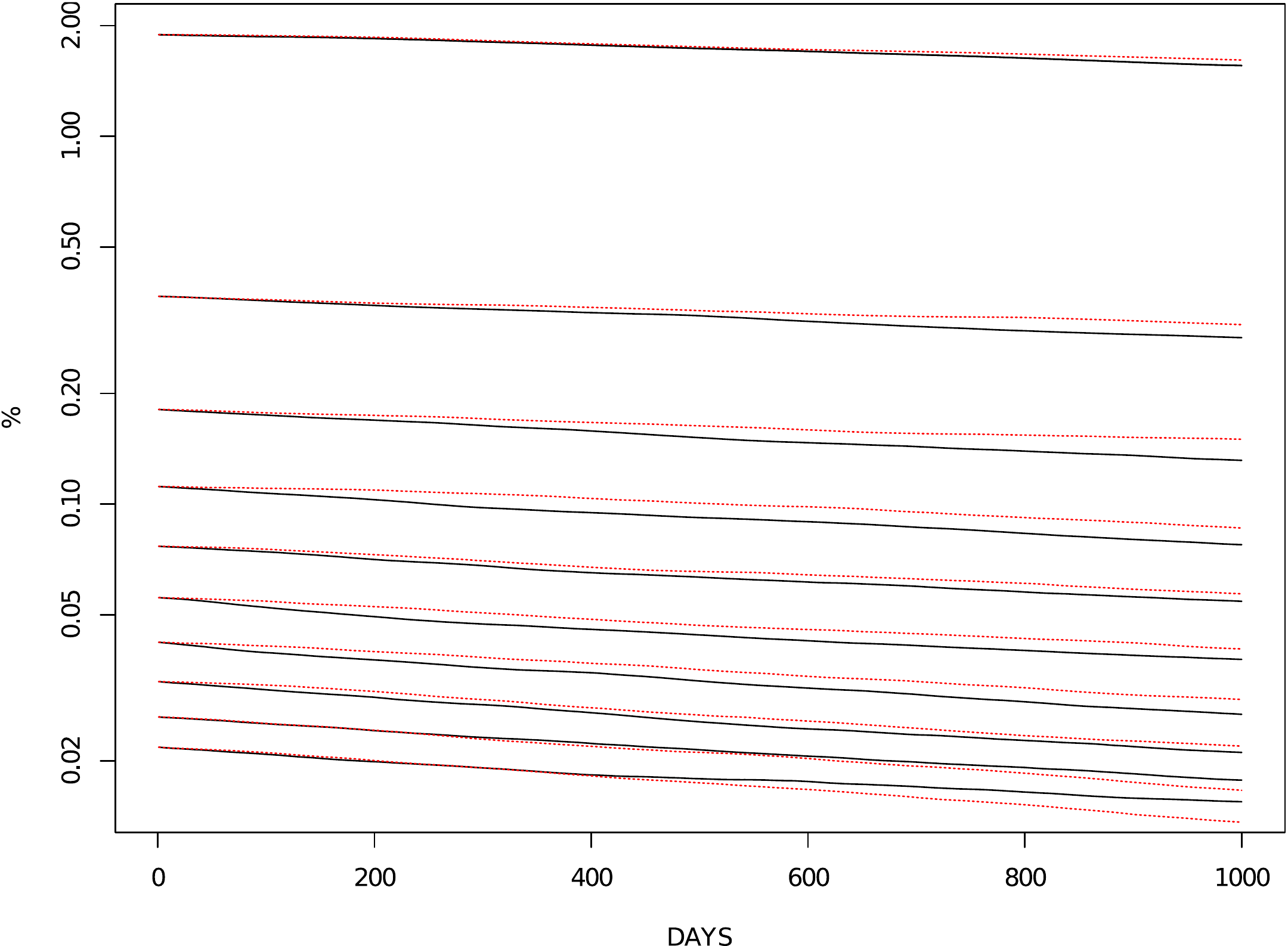}}}
\end{center}
\vspace*{-20pt}
\caption{$\m_{(k)}(0)e^{\ph_k(\t)}$ (black, solid),  $\m_{(k)}(0)e^{\widetilde{\ph}_k(\t)}$ (red, dotted): 1990--1999. On average,}\label{f4} 
\centerline{ a stock that starts at time 0 at rank $k$ with market weight $\m_{(k)}(0)$ will move to weight }
\centerline{$\m_{(k)}(0)e^{\ph_k(\t)}$ at time $\t\in[0,1000]$, or to weight $\m_{(k)}(0)e^{\widetilde\ph_k(\t)}$ in reversed time. }
\end{figure}

In order to estimate the second-order parameters, it is necessary to observe the movement of market weights forward and backward in time. To this end, we define the concept of \emph{flow} in a market.
The \emph{forward flow}  $\ph_k$ of the market at rank $k$  is defined for $\t\ge0$ by
\begin{equation*}
\ph_k(\t)\eqdef\limT{1}\intT\log\Big(\frac{\m_{p_t(k)}(t+\t)}{\m_{(k)}(t)}\Big)dt,
\end{equation*}
and the \emph{backward flow}  $\widetilde{\ph}_k$ of the market is defined by
\begin{equation*}
\widetilde{\ph}_k(\t)\eqdef\limT{1}\intT\log\Big(\frac{\widetilde\m_{p_t(k)}(t+\t)}{\widetilde\m_{(k)}(t)}\Big)dt.
\end{equation*}
In Figure~\ref{f4} we see the exponential of forward and backward flow for the largest 250 stocks in the U.S. market over the decade from 1990 to 1999. The plots  show the average exponential flow of each of the ten deciles of the top 250 stocks, with each decile comprising 25 stocks. The forward and backward flows need not be equal, and they do not appear to be equal in Figure~\ref{f4}. We see from the chart that for the largest 250 stocks the flow is downward. For the smaller stocks, we would expect the flow to be upward.

\begin{figure}[H]
\begin{center}
\vspace*{-5pt}
\hspace*{-20pt}
\scalebox{.725}{ \rotatebox{0}{\includegraphics{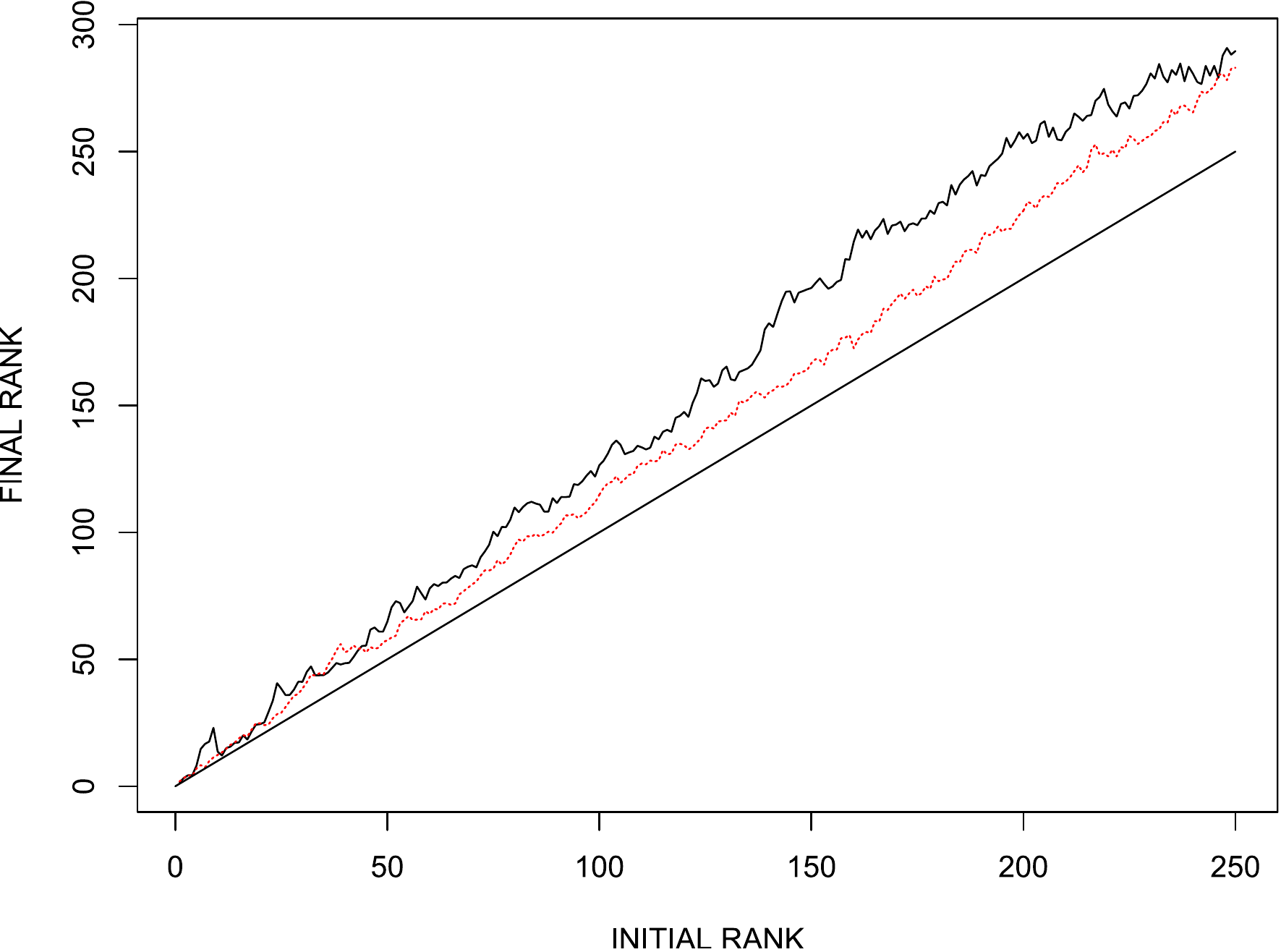}}}
\end{center}
\vspace*{-20pt}
\caption{$\RR_k(4)$ (black, solid) and $\RR_k(-4)$ (red, dotted): 1990--1999.} \label{f5}
\centerline{On average, a stock that starts a given initial rank will move  to the }
\centerline{corresponding final rank four years later, or earlier, in reversed time.}
\centerline{The straight line represents final rank $=$ initial rank.}
\end{figure}

If we follow the flow of a stock that occupies a given rank at time zero, then the expected rank of the stock will change over time according to its flow.
Suppose a stock is at rank $k$ at time 0, and 
let us estimate  its  expected rank at time $\t\in\R$ by
\begin{equation*}
\RR_k(\t)\eqdef \limT{1}\intT r_{s+\t}(p_s(k))\,ds.
\end{equation*}
In this case, $\RR_k(0)=k$, and if this rank is among the higher ranks, we would expect the flow to be negative, which would mean that for $\t>0$ we would expect that $\RR_k(\t)\le k$ and $\RR_k(-\t)\le k$. We would like to use the $\RR_k$ to estimate the $g_k$, and although $\RR_k(\t)$ need not equal $\RR_k(-\t)$, the $g_k$ generated using either one will provide estimates for the solution of \eqref{1}. Accordingly, we shall  use the average of the two, with
\begin{equation*}
\RRbar_k(\t)\eqdef\Bigg[\frac{\RR_k(\t)+\RR_k(-\t)}{2}\Bigg],
\end{equation*}
where the brackets signify the nearest integer. Values of $\RR_k(\t)$ for $k=1,\ldots,250$ and $\t=\pm4$ are shown in Figure~\ref{f5}, and the values for positive and negative $\t$ are clearly different. We have no explanation for this difference.

Figure~\ref{f4} was generated by following the market weights of stocks that occupied a given rank at a given time in the decade from January 1, 1990 to December 31, 1999. Since stocks enter and leave the market, we used only the largest 250 stocks, after eliminating any stocks that did not have a full ten-year history. The trajectories of the weights for each to the top 250 ranks were followed for 1000 days forward or backward, and were then averaged over all starting dates that would allow the full 1000 days to be used. Finally, the ranks were separated into deciles, with ranks 1--25 in the first decile, 26--50 in the second decile, and so forth. The curves in Figure~\ref{f4} represent the average trajectories for the weights of each of the ten deciles, forward and backward.

Figure~\ref{f5} was generated by following the weight trajectories used for Figure~\ref{f4} and, for each trajectory, noting the starting rank and ending rank, i.e., the rank after 1000 days (approximately four years of trading days). The final rank corresponding to the initial rank $k$ in Figure~\ref{f5} is the average ending rank for those trajectories that begin at rank $k$ at time 0. This was carried out in forward time and reversed time.

\vspace{10pt}

Let $\GG_k(\t)$ be the expected  growth rate at time $\t\in\R$ of a stock which 
occupies rank $k$ at time 0, and we shall estimate $\GG_k(\t)$ and  $\GG_k(-\t)$  from the slope of the forward and backward flow at rank $k$, so
\begin{equation}\label{4.4}
\GG_k(\t)=D_\t \ph_k(\t)\quad\text{and}\quad\GG_k(-\t)=D_\t \widetilde{\ph}_k(\t),
\end{equation}
for $\t\ge 0$, with $\GG_k(0) =\bg_k$.
We shall use the average 
\begin{equation*}
\GGbar_k(\t)\eqdef\half\big(\GG_k(\t) +\GG_k(-\t)\big)
\end{equation*}
 to estimate the rank-based growth rates $g_k$. In the data we analyzed, the derivatives in \eqref{4.4} at $\t=4$ were estimated by measuring the rate of change of the flows $\ph(\t)$ and $\widetilde\ph(\t)$ for the period from day 981 to day 1000, and then annualizing this rate.

Since for a given stock the name-based growth rate is invariant with rank, the same holds for the  average of the name-based growth rates weighted by occupation rates,
\begin{equation*}
\sumi\thhat_{ki}\g_i.
\end{equation*}
Hence,
\begin{equation}\label{4.1}
\GGbar_k(\t)\eqapprox  g_{\RRbar_k(\t)}+\sumi\thhat_{ki}\g_i,
\end{equation}
for $\t\in\R$, where
\begin{equation*}
g_{\RRbar_k(\t)}\eqdef\big(\ell+1-\RRbar_k(\t)\big)\,g_\ell+\big(\RRbar_k(\t)-
\ell\big)\,g_{\ell+1},
\end{equation*}
and $\ell$  the largest integer such that $\ell\le\RRbar_k(\t)$. If we combine \eqref{4.1} with \eqref{3.5}, we find that
\begin{equation}\label{4.2}
g_{\RRbar_k(\t)} \eqapprox g_k+\GGbar_k(\t) -\bg_k.
\end{equation}
 We can first estimate $\bg_k$, $\GGbar_k(\t)$, and $\RRbar_k(\t)$, and then use \eqref{4.2} to recursively generate the values of the rank-based growth rates $g_k$ for a subsequence of ranks of the form $k,\RRbar_k(\t),\RRbar_{\RRbar_k(\t)}(\t),\ldots$, as well as interpolated points.

Once we have estimates for the  values of  the $g_k$,  we can estimate the $\g_i$ directly by using 
\begin{equation}\label{4.3}
\g_i=\half\Big(\limT{1}\intT\big(d\log \m_i(t) - g_{r_t(i)}\,dt\big)+
\limT{1}\intT\big(d\log \mtil_i(t) - g_{r_t(i)}\,dt\big)\Big).
\end{equation}
Our second-order model for the market $X$ will then be
\begin{equation*}
d\log \Xhat_i(t) = (\g_i+g_{\rhat_t(i)})dt +\s_{\rhat_t(i)}\,dW_i(t).
\end{equation*}

\begin{figure}[H]
\begin{center}
\hspace*{-20pt}
\scalebox{.58}{ \rotatebox{90}{\includegraphics{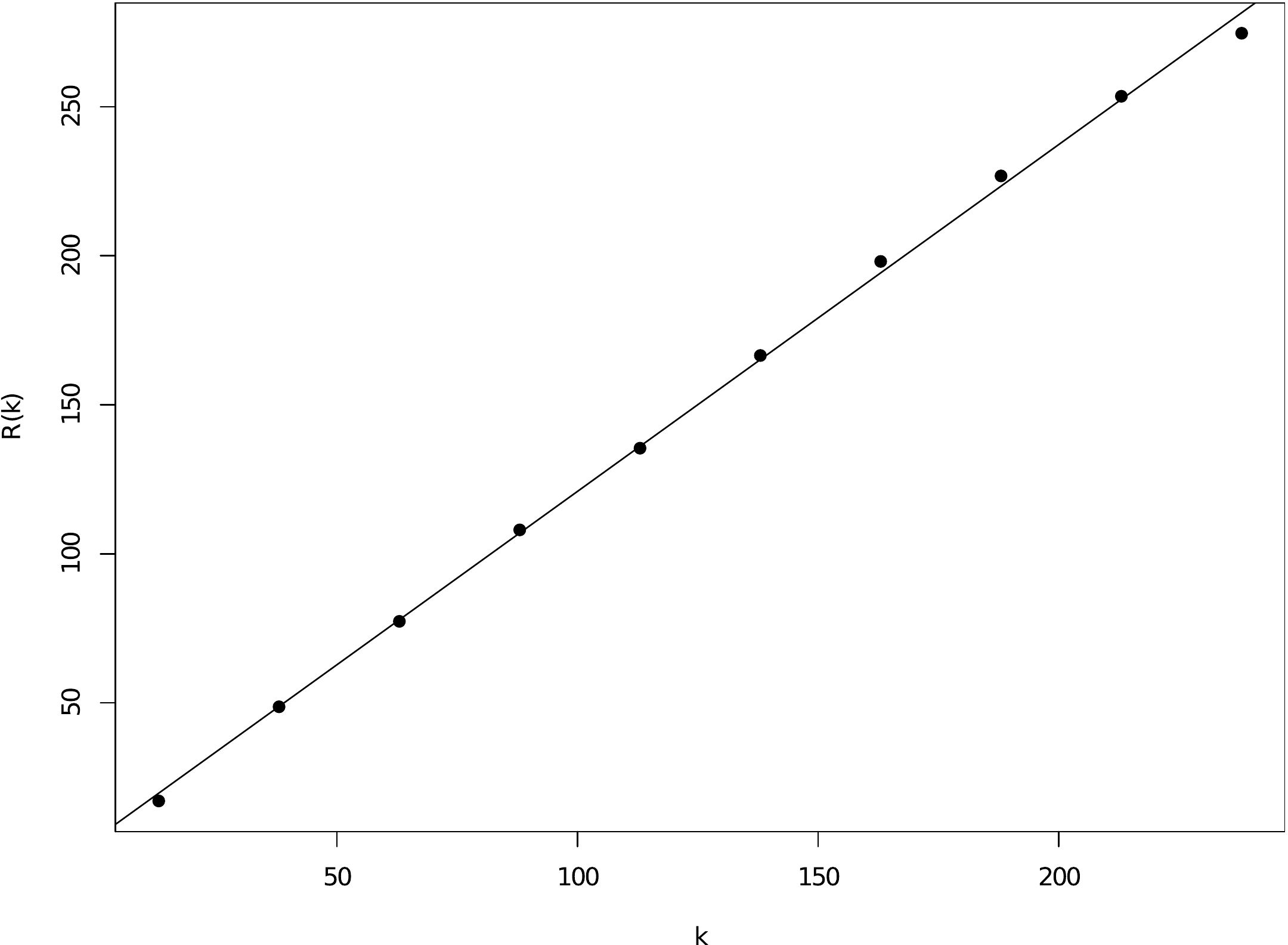}}}\end{center}
\caption{Estimated four-year forward rank $\RRbar_k(4)$ corresponding to initial rank $k$.}\label{f6}
\centerline{$\RRbar_k(4)\eqapprox 4.6+1.16k$.} 
\end{figure}

\begin{figure}[H]
\begin{center}
\hspace*{-20pt}
\scalebox{.58}{ \rotatebox{0}{\includegraphics{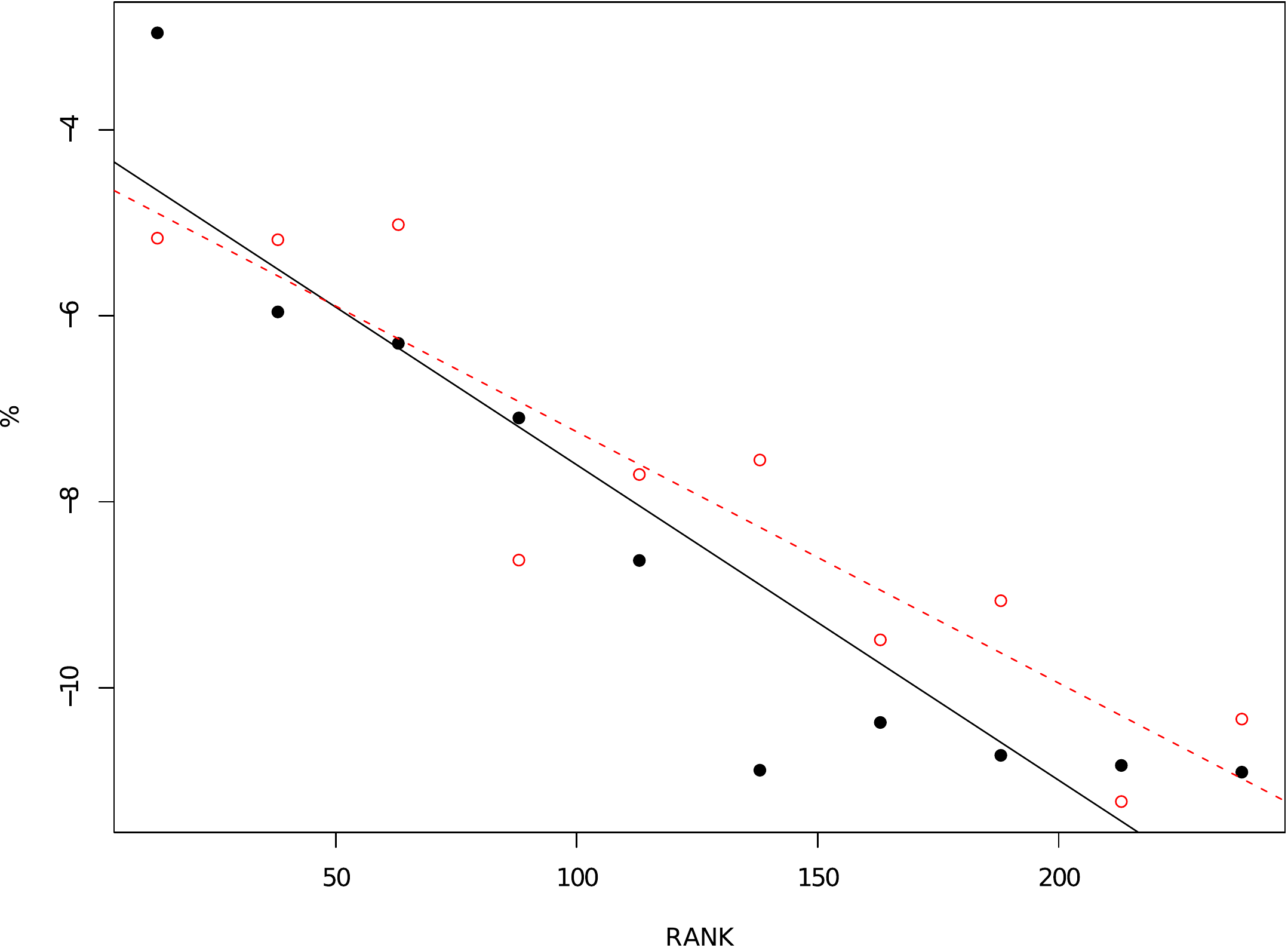}}}
\end{center}
\vspace*{-15pt}
\caption{Estimated expected growth rates $\GGbar_k(0)$ and $\GGbar_k(4)$ at rank $k$.} \label{f7}
\centerline{$\GGbar_k(0)\eqapprox-4.2-.034k$ (black, solid line; dots),}
\centerline{$\GGbar_k(4)\eqapprox-4.5-.027k$ (red, broken line; circles).}
\end{figure}

\begin{figure}[H]
\begin{center}
\hspace*{-20pt}
\scalebox{.58}{ \rotatebox{0}{\includegraphics{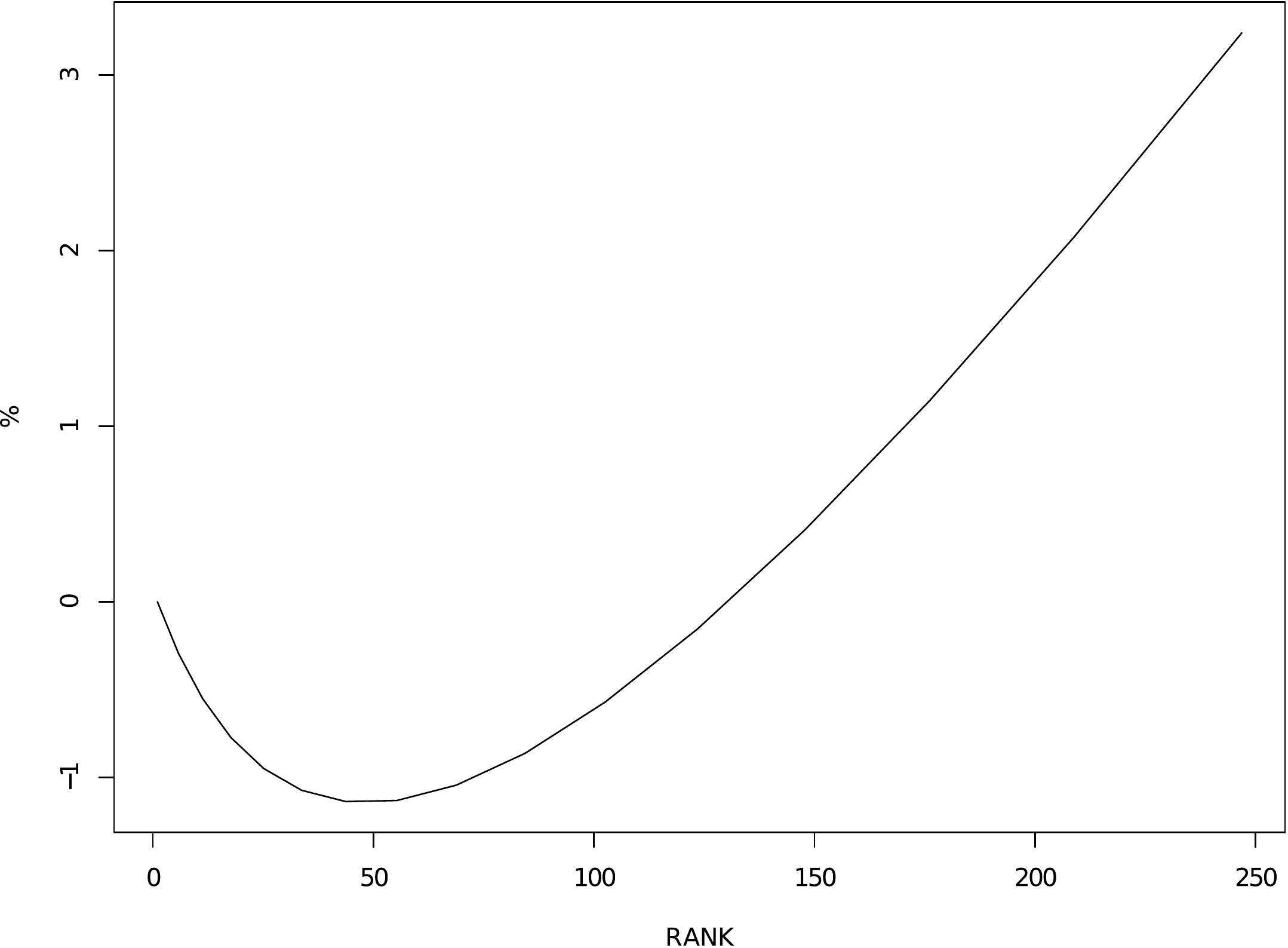}}}\end{center}
\vspace*{-15pt}
\caption{Values of $g_k$ for ranks 1 to 250, calculated recursively and interpolated from } \label{f8}
\centerline{ $g_{(4.6 + 1.16k)} \eqapprox g_k+\GGbar_k(4) -\bg_k$. The $g_k$ here are not normalized to add up to 0.}
\end{figure}

The various steps in the estimation process are shown in Figures~\ref{f6}, \ref{f7}, and \ref{f8}. In Figure~\ref{f6} we see the estimated forward rank $\RRbar_k(4)$ versus the initial rank $k$, and find that the relation is quite close to linear with
\begin{equation*}
\RRbar_k(4) \eqapprox 4.6 + 1.16k.
\end{equation*}
With this estimate, we can use \eqref{4.2} in the form
\begin{equation}\label{4.4}
g_{(4.6 + 1.16k)} \eqapprox g_k+\GGbar_k(4) -\bg_k
\end{equation}
to estimate the $g_k$ from the values of $\bg_k$ and $\GGbar_k(4)$.

The values of $\GGbar_k$ were estimated from the slopes of the flows used to generate Figure~\ref{f4} for the ten rank-decile groups of 25 stocks each from the largest 250 stocks. Linear approximations for all the ranks were generated using a least squares fit. These results appear in Figure~\ref{f7}, and the corresponding linear equations are
\begin{equation*}
\GGbar_k(0)=-4.2-.034k\quad\text{ and }\quad\GGbar_k(4)=-4.5-.027k. 
\end{equation*}
By using the values derived from these equations in \eqref{4.4} we can generate values for $g_k$ for an increasing sequence of ranks $k$. The chart in Figure~\ref{f8} shows these values with linear interpolation connecting the points to generate a continuous curve.

\bigskip

Once we have estimates for the values of the $g_k$, we can use these values along with \eqref{4.3} to estimate values for the $\g_i$ of individual stocks by name. The integrals in \eqref{4.3} were approximated by daily logarithmic relative returns taken for each of the stocks along with the values of the $g_k$. The (non-normalized) values for $\g_i$ for the decade 1990--1999 for a number of well-known stocks appear in Table~\ref{t1}. This is hardly a definitive study, so only a few stocks are included here. Moreover, in some future work, it would be desirable to have confidence intervals for these values, rather than point estimates. In that regard, probably the most promising method would use some form of jackknife estimator, with perhaps 12 pseudovalues generated by leaving out one month of the year at a time (see \citeN{Mosteller/Tukey:1977}). Probably the entire estimation process would need to be repeated for each pseudovalue.

The values in Table~\ref{t1} were estimated using combined forward and backward estimates, as in \eqref{4.3}, for the decade  1990--1999. Using only forward estimates or only backward estimates for the $\g_i$ could have produced biased estimates, since some of these companies grew considerably over that decade. For each company, the number in parentheses is the rank of the time-averaged log-weight of the stock during the decade.

While the values in Table~\ref{t1} may not be definitive, they at least appear plausible. The higher-ranked stocks have generally higher $\g$, which should help them maintain their positions at the top of the market. At this writing, Apple, AAPL,  has the highest market capitalization in the U.S. market, but in the 1990s we see that its average rank was 93, and its $\g$ is correspondingly $-1.67\%$. Hence, the estimated $\g_i$ provide no miraculous forecasts of future behavior; instead they reflect local stability consistent with the observed decade.

\begin{table}[H]
\caption{Values of $\g_i$ for various companies, 1990--1999.}\label{t1}
\begin{center}
\begin{tabular}{l>{$}r<{$}}
 Apple, AAPL (93) & -1.67\%  \\
 Coca Cola, KO (4)  & 0.26\%\\ 
Exxon, XON (3)  &0.11\% \\
  General Electric, GE (1) &0.14\%\\ 
International Business Machines, IBM (6) &\quad -0.10\%  \\ 
 Microsoft, MSFT (5)  & -0.12\% 
\end{tabular}
\end{center}
\label{default}
\end{table}

\section{Conclusion}

The purpose of first- and second-order models for stock markets is to create a rigorous backdrop for the statistical analysis of the behavior of individual stocks. Second-order models provide a more accurate and complete representation of a stock market than is possible in first-order models. The estimation of parameters for second-order models is more involved than for first-order models, and implicit methods must be used. We have proposed methods for the estimation of second-order growth rate parameters, and with these methods a more complete stock-market model is possible. Nevertheless, our techniques are rudimentary, and we believe that future research will yield significant improvements.

\bibliographystyle{plainnat}

\end{document}